\documentclass[10pt, a4paper, conference, compsocconf]{IEEEtran}
\usepackage{amsmath}
\usepackage{graphicx}
\usepackage{amsfonts}
\usepackage{amssymb}
\usepackage{alltt}
\usepackage{booktabs}
\usepackage{stmaryrd}
\usepackage{dsfont} 
\usepackage{color} 
\usepackage[standard]{ntheorem}
\usepackage[font=footnotesize]{subfig}
\usepackage{array,algorithm2e,algorithmic}
\renewcommand{\theequation}{\arabic{equation}}
\usepackage[width=18.2cm,height=25.7cm]{geometry}

\ifCLASSINFOpdf
\else
\fi
\begin{document}

\title{Ramdomness quality of CI chaotic generators. Application to Internet security}
\author{\IEEEauthorblockN{Jacques M. Bahi\IEEEauthorrefmark{1}, 
Xiaole Fang\IEEEauthorrefmark{1},
Christophe Guyeux\IEEEauthorrefmark{1} and
Qianxue Wang\IEEEauthorrefmark{1}}
\IEEEauthorblockA{\IEEEauthorrefmark{1}University of Franche-Comte\\
Computer Science Laboratory LIFC,
Besan\c con, France\\ Email:{jacques.bahi, xiaole.fang,
christophe.guyeux, qianxue.wang}@univ-fcomte.fr }}

\maketitle

\begin{abstract}
Due to the rapid development of the Internet in recent years, the need to find new tools to reinforce trust and security through the Internet has became a major concern. The discovery of new pseudo-random number generators with a strong level of security is thus becoming a hot topic, because numerous cryptosystems and data hiding schemes are directly dependent on the quality of these generators. At the conference Internet`09, we have described a generator based on chaotic iterations, which behaves chaotically as defined by Devaney. In this paper, the proposal is to improve the speed and the security of this generator, to make its use more relevant in the Internet security context. To do so, a comparative study between various generators is carried out and statistical results are given. Finally, an application in the information hiding framework is presented, to give an illustrative example of the use of such a generator in the Internet security field.
\end{abstract}
\IEEEpeerreviewmaketitle

\section{Introduction}

The development and popularity of the Internet, and its recent role in everyday life implies the need to protect data and privacy in digital world. This development has revealed new major security issues. For example, new concerns have recently appeared with the evolving of the Internet, as evoting, VoD or intellectual property protection. The pseudo-random number generators (PRNG) play an important role in all of these emerging techniques, because they are fundamental in cryptosystems and information hiding schemes. 
PRNGs are typically defined by a deterministic recurrent sequence in a finite state space, usually a finite field or ring, and an output function mapping each state to an input value. This is often either a real number in the interval $(0,1)$ or an integer in some finite range~\cite{L'ecuyer2008}. Conventionally, PRNGs based on linear congruential methods and feedback shift-registers are popular~\cite{Knuth1998}.

To use a PRNG with a large level of security is necessary to satisfy the Internet security requirements recalled above.
This level depends on the proof of theoretical properties and results of numerous statistical tests. 
Many PRNGs have been proven to be secure, following a probabilistic approach. 
However, recently, several researchers have been exploring the idea of using chaotic dynamical systems for this purpose~\cite{Falcioni2005}~\cite{Cecen2009}. 
The random-like, unpredictable dynamics of chaotic systems, their inherent determinism and simplicity of realization suggest their potential for exploitation as PRNGs.
Such generators can strongly improve security in information hiding and cryptography: due to unpredictability, the possibilities offered to an attacker to achieve his goal are drastically reduced. 
For example the keys of cryptosystems need to be unpredictable enough, making it impossible for any search optimization based on the reduction of the key space to the most probable values. 
But the number of generators claimed as chaotic, which actually have been proven to be unpredictable (as it is defined in the mathematical theory of chaos) is very small.
 
This paper extends a study initiated in~\cite{guyeux09} and~\cite{wang2009}, and tries to fill this gap. In~\cite{guyeux09}, it is proven that chaotic iterations (CIs), a suitable tool for fast computing iterative algorithms, satisfies the topological chaotic property, as it is defined by Devaney~\cite{Dev89}.
In the paper~\cite{wang2009} presented at Internet`09, the chaotic behavior of CIs is exploited in order to obtain an unpredictable PRNG, which depends on two logistic maps.
We have shown that, in addition of being chaotic, this generator can pass the NIST (National Institute of Standards and Technology of the U.S. Government) battery of tests~\cite{ANDREW2008},
widely considered as a comprehensive and stringent battery of tests for cryptographic applications.
In this paper, we have improved the speed and security of the former generator.
Chaotic properties, statistical tests and security analysis~\cite{ZHENG92008} allow us to consider that this generator has good pseudo-random characteristics and is capable to withstand attacks.
Moreover, its high linear complexity and its large key space lead to the conviction that this generator is suitable for applications in the Internet security field. 
After having presented the theoretical framework of the study and a security analysis, we will give a comparison based on statistical tests. Finally a concrete example of how to use these pseudo-random numbers for information hiding through the Internet is detailed.

The rest of this paper is organized in the following way. In Section~\ref{Basic recalls}, some basic definitions concerning chaotic iterations and PRNGs are recalled. Then, the generator based on discrete chaotic iterations is presented in Section~\ref{The generation of pseudo-random sequence}. Section~\ref{Security analysis} is devoted to its security analysis. In Section~\ref{Comparative analysis}, various tests are passed with a goal to achieve a statistical comparison between this new PRNG and other existing ones. In Section~\ref{An application example of the proposed PRNG}, a potential use of this PRNG in some Internet security field is presented, namely in information hiding. The paper ends with a conclusion and intended future work.

\section{Basic recalls}
\label{Basic recalls}

\subsection{Notations}
\begin{tabular}{@{}c@{}@{}l@{}}
$\llbracket 1;\mathsf{N} \rrbracket$ & $\rightarrow\{1,2,\hdots,N\}$ \\
$S^{n}$ & $\rightarrow$ the $n^{th}$ term of a sequence $S=(S^{1},S^{2},\hdots)$ \\
$v_{i}$ & $\rightarrow$ the $i^{th}$ component of a vector \\
        &~~~~$v=(v_{1},v_{2},\hdots, v_n)$\\
$f^{k}$ & $\rightarrow$ $k^{th}$ composition of a function $f$ \\
$\emph{strategy}$~ & $\rightarrow$ a sequence which elements belong in $%
\llbracket 1;\mathsf{N} \rrbracket $ \\
$\mathbb{S}$ & $\rightarrow$ the set of all strategies \\
$\mathbf{C}_n^k$ & $\rightarrow$ the binomial coefficient ${n \choose k} = \frac{n!}{k!(n-k)!}$\\
$\oplus$ & $\rightarrow$ bitwise exclusive or \\
$+$ & $\rightarrow$ the integer addition \\
$\ll \text{and} \gg$ & $\rightarrow$ the usual shift operators \\
$(\mathcal{X}, \text{d})$ & $\rightarrow$ a metric space  \\
\end{tabular}

\begin{tabular}{@{}c@{}@{}l@{}}

$\lfloor x \rfloor$ & $\rightarrow$ returns the highest integer smaller than $x$  \\
$n!$ & $\rightarrow$ the factorial $n!=n\times(n-1)\times\dots\times1$\\
$\mathds{N}^{\ast }$ & $\rightarrow$ the set of positive integers \{1,2,3,...\}
\end{tabular}

\subsection{Chaotic iterations}
\label{subsection:Chaotic iterations}
\begin{definition}
The set $\mathds{B}$ denoting $\{0,1\}$, let $f:\mathds{B}^{\mathsf{N}%
}\longrightarrow \mathds{B}^{\mathsf{N}}$ be an ``iteration'' function and $S\in \mathbb{S}
$ be a chaotic strategy. Then, the so-called \emph{chaotic iterations} are defined by~\cite{Robert1986}
\begin{equation}
\begin{array}{l}
x^0\in \mathds{B}^{\mathsf{N}}, \\
\forall n\in \mathds{N}^{\ast },\forall i\in \llbracket1;\mathsf{N}\rrbracket%
,x_i^n=\left\{
\begin{array}{l}
x_i^{n-1}~~~~~\text{if}~S^n\neq i \\
f(x^{n-1})_{S^n}~\text{if}~S^n=i.\end{array} \right. \end{array}
\end{equation}
\end{definition}
In other words, at the $n^{th}$ iteration, only the $S^{n}-$th cell is
\textquotedblleft iterated\textquotedblright.
Chaotic iterations generate a set of vectors (boolean vectors in this paper),
which are defined by an initial state $x^{0}$, an iteration function $f$, and a
chaotic strategy $S$.

\subsection{XORshift}
\label{XORshift}
XORshift is a category of very fast PRNGs designed by George Marsaglia~\cite{Marsaglia2003}.
It repeatedly uses the transform of exclusive or (XOR) on a number with a bit shifted version of it. The state of a XORshift generator is a vector of bits. At each step, the next state is obtained by applying a given number of XORshift operations to $w$-bit blocks in the current state, where $w = 32$ or $64$. A XORshift operation is defined as follows. Replace the $w$-bit block by a bitwise XOR of the original block, with a shifted copy of itself by $a$ positions either to the right or to the left, where $ 0 < a < w$. This Algorithm~\ref{XORshift2} has a period of $2^{32}-1=4.29\times10^9$.

\begin{algorithm}
\SetAlgoLined
\KwIn{the internal state $z$ (a 32-bits word)}
\KwOut{$y$ (a 32-bits word)}
$z\leftarrow{z\oplus{(z\ll13)}}$\;
$z\leftarrow{z\oplus{(z\gg17)}}$\;
$z\leftarrow{z\oplus{(z\ll5)}}$\;
$y\leftarrow{z}$\;
return $y$\;
\medskip
\caption{An arbitrary round of XORshift algorithm}
\label{XORshift2}
\end{algorithm}
%
%
%
%
%
%
%

\section{The new generation of CI pseudo-random sequence}
\label{The generation of pseudo-random sequence}
\subsection{Chaotic iterations as pseudo-random generator}
\subsubsection{Presentation}
The novel generator is designed by the following process. First of all, some chaotic iterations have to be done to generate a sequence $\left(x^n\right)_{n\in\mathds{N}} \in \left(\mathds{B}^\mathsf{N}\right)^\mathds{N}$ ($\mathsf{N} \in \mathds{N}^*, \mathsf{N} \geqslant 2$, $N$ is not necessarily equal to 32) of boolean vectors, which are the successive states of the iterated system. Some of these vectors will be randomly extracted and our pseudo-random bit flow will be constituted by their components. Such chaotic iterations are realized as follows. Initial state $x^0 \in \mathds{B}^\mathsf{N}$ is a boolean vector taken as a seed (see Section~\ref{algo seed}) and chaotic strategy $\left(S^n\right)_{n\in\mathds{N}}\in \llbracket 1, \mathsf{N} \rrbracket^\mathds{N}$ is 
an irregular decimation of a XORshift sequence (Section~\ref{Chaotic strategy}). The iterate function $f$ is 
the vectorial boolean negation:
$$f_0:(x_1,...,x_\mathsf{N}) \in \mathds{B}^\mathsf{N} \longmapsto (\overline{x_1},...,\overline{x_\mathsf{N}}) \in \mathds{B}^\mathsf{N}.$$
At each iteration, only the $S^i$-th component of state $X^n$ is updated, as follows: $x_i^n = x_i^{n-1}$ if $i \neq S^i$, else $x_i^n = \overline{x_i^{n-1}}$.
Finally, some $x^n$ are selected 
by a sequence $m^n$ as the pseudo-random bit sequence of our generator. The sequence 
$(m^n)_{n \in \mathds{N}} \in \mathcal{M}^\mathds{N}$ is computed from a XORshift sequence $(y^n)_{n \in \mathds{N}} \in \llbracket 0, 2^{32}-1 \rrbracket$ (see Section~\ref{algo m}). So, the 
generator returns the following values:\newline
\begin{small}
Bits:$$x_1^{m_0}x_2^{m_0}x_3^{m_0}\hdots x_\mathsf{N}^{m_0}x_1^{m_0+m_1}x_2^{m_0+m_1}\hdots x_\mathsf{N}^{m_0+m_1} x_1^{m_0+m_1+m_2}\hdots$$
or States:$$x^{m_0}x^{m_0+m_1}x^{m_0+m_1+m_2}\hdots$$
\end{small}

\subsubsection{The seed}
\label{algo seed}
%
The initial state of the system $x^0$ and the first term $y^0$ of the XORshift are seeded either by 
the current time in seconds since the Epoch, or by a number that the user inputs, as it is usually the case for every PRNG.

\subsubsection{Sequence $m$ of returned states}
\label{algo m}
The output of the sequence $(y^n)$ is uniform in $\llbracket 0, 2^{32}-1 \rrbracket$, because it is produced by a XORshift generator. However, we do not want the output of $(m^n)$ to be uniform in $\llbracket 0, N \rrbracket$, because in this case, the returns of our generator will not be uniform in $\llbracket 0, 2^{N}-1 \rrbracket$, as it is illustrated in the following example. Let us suppose that $x^0=(0,0,0)$. Then $m^0 \in \llbracket 0, 3 \rrbracket$.
\begin{itemize}
\item If $m^0=0$, then no bit will change between the first and the second output of our PRNG. Thus $x^1 = (0,0,0)$.
\item If $m^0=1$, then exactly one bit will change, which leads to three possible values for $x^1$, namely $(1,0,0)$, $(0,1,0)$, and $(0,0,1)$.
\item \emph{etc.}
\end{itemize}
As each value in $\llbracket 0, 2^3-1 \rrbracket$ must be returned with the same probability, then the values $(0,0,0)$, $(1,0,0)$, $(0,1,0)$ and $(0,0,1)$ must occur for $x^1$ with the same probability. Finally we see that, in this example, $m^0=1$ must be three times more probable than $m^0=0$.
This leads to the following general definition for $m$:
\begin{equation}
\label{Formula}
m^n = f(y^n)=
\left\{
\begin{array}{l}
0 \text{ if }0				\leqslant\frac{y^n}{2^{32}}<\frac{C^0_N}{2^N},\\
1 \text{ if }\frac{C^0_N}{2^N}		\leqslant\frac{y^n}{2^{32}}<\sum_{i=0}^1\frac{C^i_N}{2^N},\\
2 \text{ if }\sum_{i=0}^1\frac{C^i_N}{2^N}	\leqslant\frac{y^n}{2^{32}}<\sum_{i=0}^2\frac{C^i_N}{2^N},\\
\vdots~~~~~					~~\vdots~~~		    ~~~~\\
N \text{ if }\sum_{i=0}^{N-1}\frac{C^i_N}{2^N}	\leqslant\frac{y^n}{2^{32}}<1.\\
\end{array}
\right.
\end{equation}
%
%
%
%
%
%
%


\subsubsection{Chaotic strategy}
\label{Chaotic strategy}
The chaotic strategy $(S^k) \in \llbracket 1, N \rrbracket^\mathds{N}$ is generated from a second XORshift sequence $(b^k) \in \llbracket 1, N \rrbracket^\mathds{N}$. The sole difference between the sequences $S$ and $b$ is that some terms of $b$ are discarded, in such a way that: $\forall k \in \mathds{N}, (S^{M^k}, S^{M^k+1}, \hdots, S^{M^{k+1}-1})$ does not contain a same integer twice, where $M^k = \sum_{i=0}^k m^i$. Therefore, no bit will change more than once between two successive outputs of our PRNG, increasing the speed of the former generator by doing so. $S$ is said to be ``an irregular decimation'' of $b$. This decimation can be obtained by the following process.

Let $(d^1,d^2,\dots,d^N)\in \{0,1\}^N$ be a mark sequence, such that whenever $\sum_{i=1}^N d^i = m^k$, 
then $\forall i, d_i=0$ ($\forall k$, the sequence is reset when $d$ contains $m^k$ times the number 1). This mark sequence will control the XORshift sequence $b$ as follows:
\begin{itemize}
\item if $d^{b^j} \neq 1$, then $S^k=b^j$, $d^{b^j} = 1$ and $k = k+1$
\item if $d^{b^j}=1$, then $b^j$ is discarded.
\end{itemize}
For example, if $b = 142\underline{2}334 142\underline{1}\underline{1}\underline{2}\underline{2}34...$ and $m = 4241...$, then $S=1423~34~1423~4...$ Another example is given in Table~\ref{XORshift}, in which $r$ means ``reset'' and the integers which are underlined in sequence $b$ are discarded.
%

\subsection{CI(XORshift, XORshift) algorithm}
The basic design procedure of the novel generator is summed up in Algorithm~\ref{Chaotic iteration1}.
The internal state is $x$, the output state is $r$. $a$ and $b$ are those computed by the two XORshift 
generators. The value $f(a)$ is an integer, defined as in Equation~\ref{Formula}. Lastly, $\mathsf{N}$ is a constant defined by the user.
\begin{algorithm}
\SetAlgoLined
\KwIn{the internal state $x$ ($\mathsf{N}$ bits)}
\KwOut{a state $r$ of $\mathsf{N}$ bits}
\For{$i=0,\dots,N$}
{
$d_i\leftarrow{0}$\;
}
$a\leftarrow{XORshift1()}$\;
$m\leftarrow{f(a)}$\;
$k\leftarrow{m}$\;
\For{$i=0,\dots,k$}
{
$b\leftarrow{XORshift2()~mod~\mathsf{N}}$\;
$S\leftarrow{b}$\;
    \If{$d_S=0$}
    {
      $x_S\leftarrow{ \overline{x_S}}$\;
      $d_S\leftarrow{1}$\;
    }
    \ElseIf{$d_S=1$}
    {
      $k\leftarrow{ k+1}$\;
    }
}
$r\leftarrow{x}$\;
return $r$\;
\medskip
\caption{An arbitrary round of the new CI(XORshift,XORshift) generator}
\label{Chaotic iteration1}
\end{algorithm}

As a comparison, the basic design procedure of the old generator is recalled in Algorithm~\ref{Chaotic iteration2} ($a$ and $b$ are computed by Logistic maps, $\mathsf{N}$ and $c\geqslant 3\mathsf{N}$ are constants defined by the user). See~\cite{wang2009} for further informations.

\begin{algorithm}
\SetAlgoLined
\KwIn{the internal state $x$ ($\mathsf{N}$ bits)}
\KwOut{a state $r$ of $\mathsf{N}$ bits}
$a\leftarrow{Logistic map1()}$\;
    \If{$a>0.5$}
      {
      $d\leftarrow 1$
      }
    \Else
      {
      $d\leftarrow 0$
      }

$m\leftarrow{d+c}$\;
\For{$i=0,\dots,m$}
{
$b\leftarrow{Logistic map2()}$\;
$S\leftarrow{100000b~mod~\mathsf{N}}$\;
$x_S\leftarrow{ \overline{x_S}}$\;
}
$r\leftarrow{x}$\;
return $r$\;
\medskip
\caption{An arbitrary round of the old PRNG}
\label{Chaotic iteration2}
\end{algorithm}

\subsection{Illustrative example}
In this example, $\mathsf{N} = 4$ is chosen for easy understanding.
The initial state of the system $x^0$ can be seeded by the decimal part $t$ of the current time. 
For example, if the current time in seconds since the Epoch is 1237632934.484088, 
so $t = 484088$, then $x^0 = t \text{ (mod 16)}$ in binary digits, \emph{i.e.}, $x^0 = ( 0, 1, 0, 0)$. 

To compute $m$ sequence, Equation~\ref{m1 fuction} can be adapted to this example as follows:
\begin{equation}
\label{m1 fuction}
m^n=f(y^n)=
\left\{
\begin{array}{llccccc}
0 & \text{ if }&0				&\leqslant&\frac{y^n}{2^{32}}&<&\frac{1}{16},\\
1 & \text{ if }&\frac{1}{16}			&\leqslant&\frac{y^n}{2^{32}}&<&\frac{5}{16} ,\\
2 & \text{ if }&\frac{5}{16}			&\leqslant&\frac{y^n}{2^{32}}&<&\frac{11}{16},\\
3 & \text{ if }&\frac{11}{16}			&\leqslant&\frac{y^n}{2^{32}}&<&\frac{15}{16},\\
4 & \text{ if }&\frac{15}{16}			&\leqslant&\frac{y^n}{2^{32}}&<&1,\\
\end{array}
\right.
\end{equation}

\noindent where $y$ is generated by XORshift seeded with the current time. We can see that the probabilities of occurrences of $m=0$, $m=1$, $m=2$, $m=3$, $m=4$, are $\frac{1}{16}$, $\frac{4}{16}$, $\frac{6}{16}$, $\frac{4}{16}$, $\frac{1}{16}$, respectively. This $m$ determines what will be the next output $x$. For instance, 
\begin{itemize}
\item If $m=0$, the following $x$ will be $( 0, 1, 0, 0)$.
\item If $m=1$, the following $x$ can be $( 1, 1, 0, 0)$, $( 0, 0, 0, 0)$, $( 0, 1, 1, 0)$ or $( 0, 1, 0, 1)$.
\item If $m=2$, the following $x$ can be $( 1, 0, 0, 0)$, $( 1, 1, 1, 0)$, $( 1, 1, 0, 1)$, $( 0, 0, 1, 0)$, $( 0, 0, 0, 1)$ or $( 0, 1, 1, 1)$.
\item If $m=3$, the following $x$ can be $( 0, 0, 1, 1)$, $( 1, 1, 1, 1)$, $( 1, 0, 0, 1)$ or $( 1, 0, 1, 0)$.
\item If $m=4$, the following $x$ will be $( 1, 0, 1, 1)$.
\end{itemize}

In this simulation, $m = 0, 4, 2, 2, 3, 4, 1, 1, 2, 3, 0, 1, 4,...$ Additionally, $b$ is computed with a XORshift generator too, but with another seed. We have found $b = 1, 4, 2, 2, 3, 3, 4, 1, 1, 4, 3, 2, 1,...$

Chaotic iterations are made with initial state $x^0$, vectorial logical negation $f_0$ and 
strategy $S$. The result is presented in Table~\ref{table application example}. Let us recall that sequence $m$ gives the states $x^n$ to return, which are here $x^0, x^{0+4}, x^{0+4+2}, \hdots$ So, in this example, the output of the generator is: 10100111101111110011... or 4,4,11,8,1... 

\begin{tiny}
\begin{table*}[!t]
\centering
\begin{tabular}{|c|cc|cccccc|ccc|cccc|}
\hline
$m$ &0 & &4 & & & & & &2& &&2&&  &  \\ \hline
$k$ &0 & &4 & & &$+1$ & & &2& &&2&$+1$&  &  \\ \hline
$b$  &  & &1 &4&2&\underline{2}       &3& &3&4&&1&\underline{1}      &4&\\ \hline
$d$  &r  & &r~(1,0,0,0)&(1,0,0,1) &(1,1,0,1)& &(1,1,1,1,)&&r~(0,0,1,0) &(0,0,1,1) &&r~(1,0,0,0) & &(1,0,0,1)  &  \\ \hline
$S$  &  & &1 &4&2&        &3& &3&4&&1& &4 &  \\ \hline
$x^{0}$ &  &$x^{0}$ & & &  
&  & &$x^{4}$ & & &   
$x^{6}$& & &&$x^{8}$  \\
0 & &0 &$\xrightarrow{1} 1$ & &
 & &   &1   & & &
1 &$\xrightarrow{1} 0$ & & & 0\\
1 &  &1 &   &   &
$\xrightarrow{2} 0$ & & &0 & & &
0 & &  &&0\\
0 & &0 & & &
 & &$\xrightarrow{3} 1$ &1 &$\xrightarrow{3} 0$ & &
0 &   & & &0  \\
0 & &0  & &$\xrightarrow{4} 1$ &
 & & &1 & &$\xrightarrow{4} 0$ &
0 & & &$\xrightarrow{4} 1$&1 \\
\hline
\end{tabular}\\
\vspace{0.5cm}
Binary Output: $x_1^{0}x_2^{0}x_3^{0}x_4^{0}x_1^{0}x_2^{0}x_3^{0}x_4^{0}x_1^{4}x_2^{4}x_3^{4}x_4^{4}x_1^{6}x_2^{6}... = 01000100101110000001...$\\
Integer Output:
$x^{0},x^{0},x^{4},x^{6},x^{8}... = 4,4,11,8,1...$
\caption{Application example}
\label{table application example}
\end{table*}
\end{tiny}

\section{Security analysis}
\label{Security analysis}
\subsection{Key space}
The PRNG proposed in this paper is based on discrete chaotic iterations. It has an initial 
value $x^0\in \mathds{B}^{\mathsf{N}}$. Considering this set of initial values alone, the key space size 
is equal to $2^\mathsf{N}$. In addition, this new generator combines digits of two other PRNGs. We used two different XORshifts here. Let $k$ be the key space of XORshift. So the total key space 
size is close to $2^\mathsf{N}\cdot k^2$. Lastly, the impact of Equation~\ref{Formula} must be 
taken into account.
This leads to conclude that the key space size is large enough to withstand 
attacks.

\subsection{Devaney's chaos property}
Generally, the quality of a PRNG depends, to a large extent, on the following criteria: randomness, uniformity, independence, storage efficiency, and reproducibility. A chaotic sequence may satisfy these requirements and also other chaotic properties, as ergodicity, entropy, and expansivity. A chaotic sequence is extremely sensitive to the initial conditions. That is, even a minute difference in the initial state of the system can lead to enormous differences in the final state, even over fairly small timescales. Therefore, chaotic sequence fits the requirements of pseudo-random sequence well. Contrary to XORshift, our generator possesses these chaotic properties~\cite{guyeux09},\cite{wang2009}.
However, despite a large number of papers published in the field of chaos-based pseudo-random generators, the impact of this research is rather marginal. This is due to the following reasons: almost all PRNG algorithms using chaos are based on dynamical systems defined on continuous sets (\emph{e.g.}, the set of real numbers). So these generators are usually slow, requiring considerably more storage space and lose their chaotic properties during computations. These major problems restrict their use as generators~\cite{Kocarev2001}.\newline
In this paper we don't simply integrate chaotic maps hoping that the implemented algorithm remains chaotic. Indeed, the PRNG we conceive is just discrete chaotic iterations and we have proven in \cite{guyeux09} that these iterations produce a topological chaos as defined by Devaney: they are regular, transitive, and sensitive to initial conditions. This famous definition of a chaotic behavior for a dynamical system implies unpredictability, mixture, sensitivity, and uniform repartition. Moreover, as only integers are manipulated in discrete chaotic iterations, the chaotic behavior of the system is preserved during computations, and these computations are fast.

\subsection{Key sensitivity}
\begin{figure}
\centering
\includegraphics[width=3.5in]{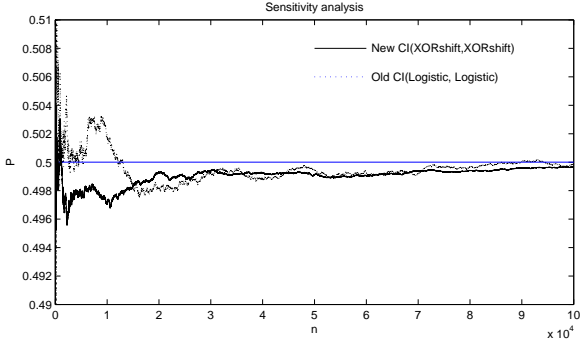}
\DeclareGraphicsExtensions.
\caption{Sensitivity analysis}
\label{Sensitivity analysis}
\end{figure}

As a consequence of its chaotic property, this PRNG is highly sensitive to the initial conditions. To illustrate this property, several initial values are put into the chaotic system. Let $H$ be the number 
of differences between the sequences obtained in this way. Suppose $n$ is the length of these 
sequences. Then the variance ratio $P$, defined by $P = H / n$, is computed. The results are 
shown in Figure~\ref{Sensitivity analysis} ($x$ axis is sequence lengths, $y$ axis is variance ratio $P$). For the two PRNGs, variance 
ratios approach $0.50$, which indicates that the system is extremely sensitive to the initial 
conditions.

\section{Statistical analysis}
\label{Comparative analysis}
\subsection{Basic usual tests}

\subsubsection{Comparative test parameters}
In this section, five well-known statistical tests~\cite{Menezes1997} are used as  comparison tools. They encompass frequency and autocorrelation tests. In what follows, $s = s^0,s^1,s^2,\dots , s^{n-1}$ denotes a binary sequence of length $n$. The question is to determine whether this sequence possesses some specific characteristics that a truly random sequence would be likely to exhibit. The tests are introduced in this subsection and results are given in the next one.

\paragraph{Frequency test (monobit test)}
The purpose of this test is to check if the numbers of 0's and 1's are approximately equal in $s$, as it would be expected for a random sequence. Let $n_0, n_1$ denote these numbers. The statistic used here is $X_1=\frac{(n_0-n_1)^2}{n}$, which approximately follows a $\chi^2$ distribution with one degree of freedom when $n\geqslant 10^7$.

\paragraph{Serial test (2-bit test)}
The purpose of this test is to determine if the number of occurrences of 00, 01, 10 and 11 as subsequences of $s$ are approximately the same. Let $n_{00} , n_{01} ,n_{10}$, and $n_{11}$ denote the number of occurrences of $00, 01, 10$, and $11$ respectively. Note that $n_{00} + n_{01} + n_{10} + n_{11} = n-1$ since the subsequences are allowed to overlap. The
statistic used here is:\newline
$X_2=\frac{4}{n-1}(n_{00}^2+n_{01}^2+n_{10}^2+n_{11}^2)-\frac{2}{n}(n_0^2+n_1^2)+1,$
\noindent which approximately follows a $\chi^2$ distribution with 2 degrees of freedom if $n\geqslant 21$.

\paragraph{Poker test}
The poker test studies if each pattern of length $m$ (without overlapping) appears the same number of times in $s$. Let $\lfloor \frac{n}{m} \rfloor\geqslant 5 \times 2^m$ and $k= \lfloor \frac{n}{m} \rfloor $. Divide the sequence $s$ into $k$ non-overlapping parts, each of length $m$. Let $n_i$ be the number of occurrences of the $i^{th}$ type of sequence of length $m$, where $1 \leqslant i \leqslant 2^m$. The statistic used is 
\begin{equation*}
X_3=\dfrac{2^m}{k}\left(\displaystyle{\sum^{2^m}_{i=1}n^2_i}\right)-k,
\end{equation*}
which approximately follows a $\chi^2$ distribution with $2^m-1$ degrees of freedom. Note that the poker test is a generalization of the frequency test: setting $m = 1$ in the poker test yields the frequency test.

\paragraph{Runs test}
The purpose of the runs test is to figure out whether the number of runs of various lengths in the sequence $s$ is as expected, for a random sequence. A run is defined as a pattern of all zeros or all ones, a block is a run of ones, and a gap is a run of zeros. The expected number of gaps (or blocks) of length $i$ in a random sequence of length $n$ is $e_i = \frac{n-i+3}{2^{i+2}}$. Let $k$ be equal to the largest integer $i$ such that $e_i \geqslant 5$. Let
$B_i , G_i$ be the number of blocks and gaps of length $i$ in $s$, for each $i \in \llbracket 1, k\rrbracket$. The statistic used here will then be:
\begin{equation*}
\displaystyle{X_4=\sum^k_{i=1}\frac{(B_i-e_i)^2}{e_i}+\sum^k_{i=1}\frac{(G_i-e_i)^2}{e_i}},
\end{equation*}
\noindent which approximately follows a $\chi^2$ distribution with $2k - 2$ degrees of freedom.

\paragraph{Autocorrelation test}
The purpose of this test is to check for coincidences between the sequence $s$ and (non-cyclic) shifted versions of it. Let $d$ be a fixed integer, $ 1 \leqslant d \leqslant \lfloor n/2 \rfloor$. The  $A(d) = \sum_{i=0}^{n-d-1} s_i\oplus s_{i+d}$ is the amount of bits not equal between the sequence and itself displaced by $d$ bits. The statistic used is:
$X_5=2(A(d)-\frac{n-d}{2})/\sqrt{n-d}$,
which approximately follows a normal distribution $N(0, 1)$ if $n-d \geqslant 10$. Since small values of $A(d)$ are as unexpected as large values, a two-sided test should be used.

\subsubsection{Comparison}
\begin{figure}
\centering
\includegraphics[width=3.7in]{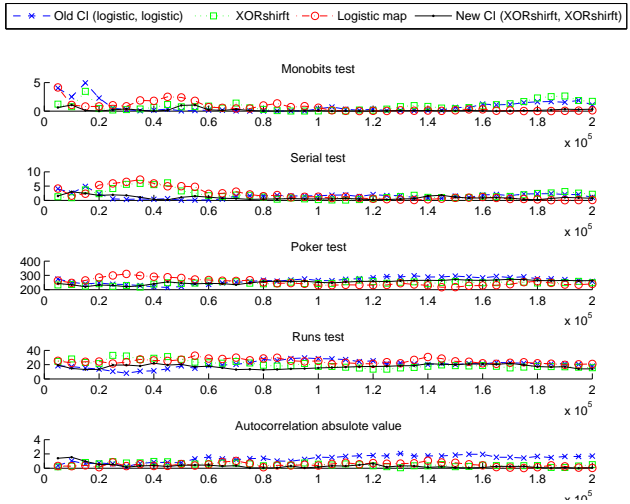}
\caption{Comparison through various well-known tests}
\label{Comparison1}
\end{figure}

\begin{table*}[!t]
\renewcommand{\arraystretch}{1.3}
\caption{Comparison with Old CI(Logistic, Logistic) for a $2 \times 10^5$ bits sequence}
\label{Comparison2}
\centering
  \begin{tabular}{ccccccc}
    \hline
Method & Monobit & Serial & Poker & Runs & Autocorrelation & Time  \\ \hline
Logistic map &0.1280&0.1302&240.2893&26.5667&0.0373&0.965s \\
XORshift &1.7053&2.1466&248.9318&18.0087&-0.5009&0.096s \\
Old CI(Logistic, Logistic) &1.0765&1.0796&258.1069&20.9272&-1.6994&0.389s \\
New CI(XORshift,XORshift) &0.3328&0.7441&262.8173&16.7877&-0.0805&0.197s\\
    \hline
  \end{tabular}
\end{table*}
We show in Table~\ref{Comparison2} a comparison between our new generator CI(XORshift, XORshift), its old version denoted Old CI(Logistic, Logistic), a PRNG based on logistic map, and a simple XORshift. Time (in seconds) is related to the duration needed by each algorithm to generate a $2 \times 10^5$ bits long sequence. The test has been conducted using the same computer and compiler with the same optimization settings for both algorithms, in order to make the test as fair as possible. Similar results have been achieved for different sequence lengths (see Figure \ref{Comparison1}).
The results confirm that the proposed generator is a lot faster than the old one, while the statistical results are better for most of the parameters, leading to the conclusion that the new PRNG is more secure than the old one. Although the logistic map also has good results, it is too slow to be implemented in Internet applications.

\subsection{NIST statistical test suite}
Among the numerous standard tests for pseudo-randomness, a convincing way to prove the quality of the produced sequences is to confront them with the NIST (National Institute of  Standards and Technology) Statistical Test Suite SP 800-22, 
released by the Information Technology Laboratory in August 25, 2008. This package of 15 tests was developed to measure the randomness of (arbitrarily long) binary sequences produced by either hardware or software based cryptographic (pseudo-)random number generators. These tests focus on a variety of different types of non-randomness that could occur in such sequences.

In our experiments, 100 sequences (s = 100) of 1,000,000 bits are generated and tested. If the value $\mathbb{P}_T$ of any test is smaller than 0.0001, the sequences are considered to not be good enough and the generator is unsuitable. Table~\ref{The passing rate} shows $\mathbb{P}_T$ of the sequences based on discrete chaotic iterations using different schemes. If there are at least two statistical values in a test, this test is marked with an asterisk and the average value is computed to characterize the statistical values. 
We can conclude from Table \ref{The passing rate} that both the old generator and CI(XORshift, XORshift) have successfully passed the NIST statistical test suite.
\begin{table}[!t]
\renewcommand{\arraystretch}{1.3}
\caption{SP 800-22 test results ($\mathbb{P}_T$)}
\label{The passing rate}
\centering
  \begin{tabular}{|l||c|c|}
    \hline
Method & Old CI & New CI  \\ \hline\hline

Frequency (Monobit) Test 			&0.595549&0.474986 \\ \hline
Frequency Test within a Block  			&0.554420&0.897763  \\ \hline
Runs Test 					&0.455937&0.816537 \\ \hline
Longest Run of Ones in a Block Test 		&0.016717&0.798139   \\ \hline
Binary Matrix Rank Test 			&0.616305&0.262249  \\ \hline
Discrete Fourier Transform (Spectral) Test	&0.000190&0.007160   \\ \hline
Non-overlapping Template Matching Test* 	&0.532252&0.449916 \\ \hline
Overlapping Template Matching Test   		&0.334538&0.514124  \\ \hline
Maurer’s “Universal Statistical” Test   	&0.032923&0.678686  \\ \hline
Linear Complexity Test  			&0.401199&0.657933    \\ \hline
Serial Test* (m=10) 				&0.013396&0.425346  \\ \hline
Approximate Entropy Test (m=10) 		&0.137282&0.637119  \\ \hline
Cumulative Sums (Cusum) Test* 			&0.046464&0.279680\\ \hline
Random Excursions Test* 			&0.503622&0.287409   \\ \hline
Random Excursions Variant Test* 		&0.347772&0.486686    \\ \hline
Success & 15/15 & 15/15 \\ \hline
    \hline
  \end{tabular}
\end{table}

\section{Application example in digital watermarking}
\label{An application example of the proposed PRNG}

\begin{figure}[!t]
\centering
\subfloat [The original image]{\includegraphics[scale=0.18]{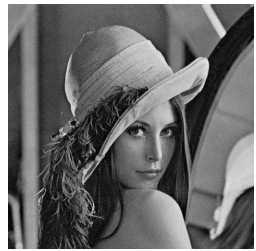}}
\hfil
\subfloat [The watermark]{\includegraphics[scale=0.4]{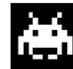}%
}
\caption{Original images}
\label{Original images}
\end{figure}

In this section, an application example is given in the field of digital watermarking: a watermark is encrypted and embedded into a cover image using the scheme presented in~\cite{guyeux10ter} and  CI(XORshift, XORshift). The carrier image is the well-known Lena, which is a 256 grayscale image, and the watermark is the $64\times 64$ pixels binary image depicted in Figure~\ref{Original images}.

\begin{figure}[!t]
\centering
\subfloat [Differences with the original]{\includegraphics[scale=0.35]{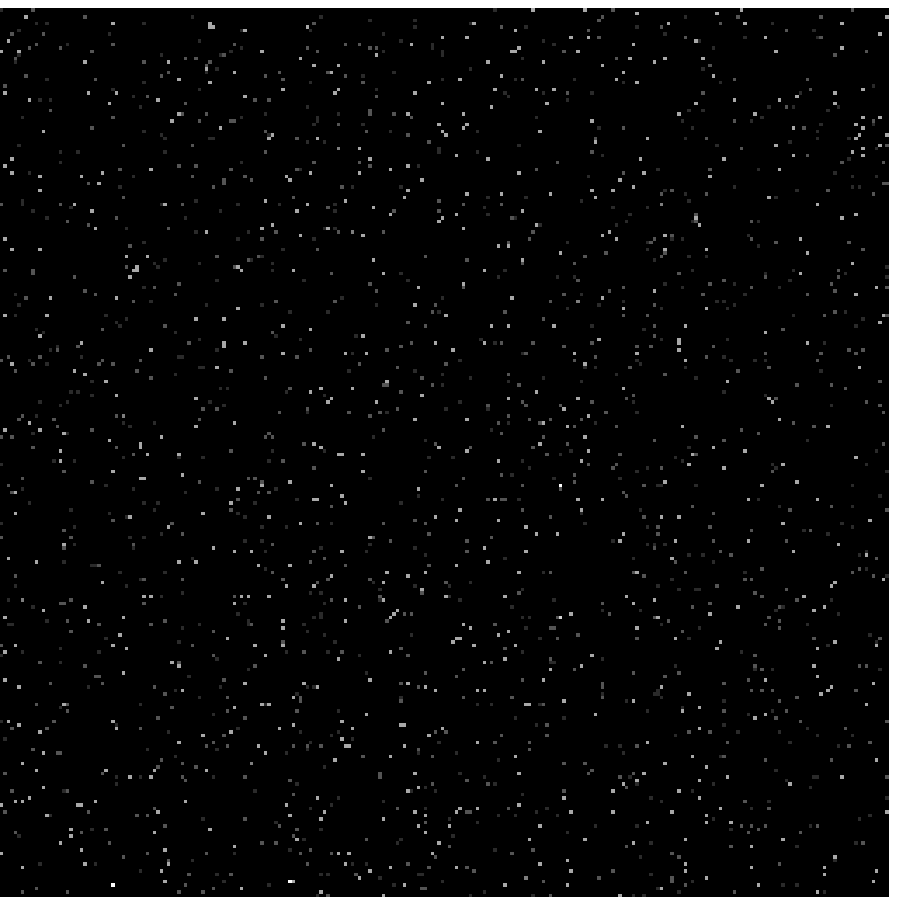}%
}
\hfil
\subfloat [The encrypted watermark]{\includegraphics[scale=0.4]{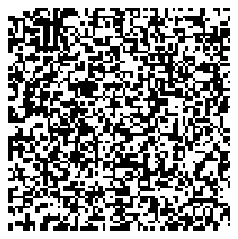}%
}
\caption{Encrypted watermark and differences}
\label{Encrypted watermark and differences}
\end{figure}

The watermark is encrypted by using chaotic iterations: the initial state $x^{0}$ is the watermark, considered as a boolean vector, the iteration function is the vectorial logical negation, and the chaotic strategy $(S^{k})_{k\in \mathds{N}}$ is defined with CI(XORshift, XORshift), where initial parameters constitute the secret key and $N=64$. Thus, the encrypted watermark is the last boolean vector generated by these chaotic iterations. An example of such an encryption is given in Figure~\ref{Encrypted watermark and differences}.

Let $L$ be the $256^3$ booleans vector constituted by the three last bits of each pixel of Lena and $U^k$ defined by:
\begin{equation}
\left\{ 
\begin{array}{lll}
U^{0} & = & S^{0} \\ 
U^{n+1} & = & S^{n+1}+2\times U^{n}+n ~ [mod ~ 256^3]%
\end{array}%
\right.
\end{equation}
The watermarked Lena $I_w$ is obtained from the original Lena, whose three last bits are replaced by the result of $64^2$ chaotic iterations with initial state $L$ and strategy $U$ (see Figure~\ref{Encrypted watermark and differences}).

The extraction of the watermark can be obtained in the same way. Remark that the map $\theta \mapsto 2\theta $ of the torus, which is the famous dyadic transformation (a well-known example of topological chaos~\cite{Dev89}), has been chosen to make $(U^{k})_{k \leqslant 64^2}$ highly sensitive to the strategy. As a consequence, $(U^{k})_{k \leqslant 64^2}$ is highly sensitive to the alteration of the image: any significant modification of the watermarked image will lead to a completely different extracted watermark, thus giving a way to authenticate media through the Internet.

\section{Conclusion and future work}
\label{Conclusions and Future Work}

In this paper, the pseudo-random generator proposed in \cite{wang2009} has been improved. By using XORshift instead of logistic map and due to a rewrite of the way to generate strategies, the generator based on chaotic iterations works faster and is more secure. The speed and randomness of this new PRNG has been compared to its former version, to XORshift, and to a generator based on logistic map. This comparison shows that CI(XORshift, XORshift) offers a sufficient speed and level of security for a whole range of Internet usages as cryptography and data hiding. 

In future work, we will continue to try to improve the speed and security of this PRNG, by exploring new strategies and iteration functions. Its chaotic behavior will be deepened by using the various tools provided by the mathematical theory of chaos. New statistical tests will be used to compare this PRNG to existing ones. Additionally a probabilistic study of its security will be done. Lastly, new applications in computer science will be proposed, especially in the Internet security field.
\bibliographystyle{plain}
\bibliography{Generating_good_chaotic_random_numbers.bib}

\end{document}